\documentclass{pasj00}
\draft
\begin{document}
\SetRunningHead{N. Isobe et al.}
{Bright X-ray flares from Mrk 421, detected with MAXI}

\Received{yyyy/mm/dd}
\Accepted{yyyy/mm/dd}

\title{Bright X-ray flares from the BL Lac object Mrk 421,
detected with MAXI in 2010 January and February}

\author{
Naoki Isobe\altaffilmark{1}, 
Kousuke Sugimori\altaffilmark{2}, 
Nobuyuki Kawai\altaffilmark{2,3}, 
Yoshihiro Ueda\altaffilmark{1}, 
Hitoshi    Negoro\altaffilmark{4}, \\
Mutsumi    Sugizaki\altaffilmark{3},
Masaru Matsuoka\altaffilmark{3,5},

Arata      Daikyuji\altaffilmark{6},
Satoshi    Eguchi\altaffilmark{1}, 
Kazuo      Hiroi\altaffilmark{1}, \\
Masaki     Ishikawa\altaffilmark{5},
Ryoji      Ishiwata\altaffilmark{4},
Kazuyoshi  Kawasaki\altaffilmark{5},
Masashi    Kimura\altaffilmark{7},  
Mitsuhiro  Kohama\altaffilmark{3,5}, \\
Tatehiro   Mihara\altaffilmark{3},  
Sho        Miyoshi\altaffilmark{4}, 
Mikio      Morii\altaffilmark{2}, 
Yujin E.   Nakagawa\altaffilmark{3}, 
Satoshi    Nakahira\altaffilmark{8},\\ 
Motoki     Nakajima\altaffilmark{9}, 
Hiroshi    Ozawa\altaffilmark{4}, 
Tetsuya    Sootome\altaffilmark{3},
Motoko     Suzuki\altaffilmark{3}, 
Hiroshi    Tomida\altaffilmark{5}, \\
Hiroshi    Tsunemi\altaffilmark{7}, 
Shiro      Ueno\altaffilmark{5}, 
Takayuki   Yamamoto\altaffilmark{3}, 
Kazutaka   Yamaoka\altaffilmark{8}, 
Atsumasa   Yoshida\altaffilmark{8}, 
\\ and  the MAXI team}

\altaffiltext{1}{Department of Astronomy, Kyoto University, Oiwake-cho, 
  Sakyo-ku, Kyoto 606-8502, Japan}
\email{n-isobe@kusastro.kyoto-u.ac.jp}
\altaffiltext{2}{Department of Physics, Tokyo Institute of Technology, 2-12-1 Ookayama, \\
  Meguro-ku, Tokyo 152-8551, Japan}
\altaffiltext{3}{Coordinated Space Observation and Experiment Research Group, \\
  Institute of Physical and Chemical Research (RIKEN), 
  2-1 Hirosawa, Wako, Saitama 351-0198, Japan}
\altaffiltext{4}{Department of Physics, Nihon University, 
  1-8-14 Kanda-Surugadai, Chiyoda-ku, \\
                 Tokyo 101-8308, Japan}
\altaffiltext{5}{ISS Science Project Office, 
  Institute of Space and Astronautical Science (ISAS), \\
  Japan Aerospace Exploration Agency (JAXA), 2-1-1 Sengen, 
  Tsukuba, Ibaraki 305-8505, Japan}
\altaffiltext{6}{Department of applied physics, University of Miyazaki, 
  1-1 Gakuen Kibanadai-nishi, \\ 
  Miyazaki, Miyazaki 889-2192, Japan}
\altaffiltext{7}{Department of Earth and Space Science, Osaka University, 
  1-1 Machikaneyama, Toyonaka, \\ 
  Osaka 560-0043, Japan}
\altaffiltext{8}{Department of Physics and Mathematics, Aoyama Gakuin University, \\ 
  5-10-1 Fuchinobe, Chuo-ku, Sagamihara, Kanagawa 252-5258, Japan}
\altaffiltext{9}{School of Dentistry at Matsudo, Nihon University, 
  2-870-1 Sakaecho-nishi, Matsudo, \\ 
  Chiba 101-8308, Japan}


\maketitle

\begin{abstract}
Strong X-ray flares from the blazar Mrk 421 were 
detected in 2010 January and February through the 7 month monitoring 
with the MAXI GSC.
The maximum 2 -- 10 keV flux in the January and February flares 
was measured as $120 \pm 10$ mCrab and $164 \pm 17 $ mCrab respectively; 
the latter is the highest among those reported from the object. 
A comparison of the MAXI and Swift BAT data suggests 
a convex X-ray spectrum 
with an approximated photon index of $\Gamma \gtrsim 2$. 
This spectrum is consistent with a picture that MAXI 
is observing near the synchrotron peak frequency. 
The source exhibited a spectral variation during these flares,
slightly different from those in the previous observations,
in which the positive correlation between the flux and hardness 
was widely reported. 
By equating the halving decay timescale in the January flare, 
$t_{\rm d} \sim 2.5 \times 10^{4}$ s, 
to the synchrotron cooling time, the magnetic field was evaluated 
as $B \sim 4.5 \times 10^{-2} $ G $(\delta/10)^{-1/3}$. 
Assuming that the light crossing time of the emission region 
is shorter than the doubling rise time, $t_{\rm r} \lesssim 2 \times 10^{4}$ s,
the region size was roughly estimated 
as $ R < 6 \times 10^{15}$ cm  $(\delta/10)$.
These are consistent with the values previously reported. 
For the February flare, 
the rise time, $t_{\rm r} < 1.3 \times 10^{5}$ s, gives 
a loose upper limit on the size as $ R < 4 \times 10^{16}$ cm $(\delta/10)$,
although the longer decay time $t_{\rm d} \sim 1.4 \times 10^{5}$ s, 
indicates $B \sim 1.5 \times 10^{-2} $ G $(\delta/10)^{-1/3}$, 
which is weaker than the previous results.
This could be reconciled by invoking a scenario 
that this flare is a superposition of unresolved events 
with a shorter timescale.
\color{black} 
\end{abstract}  

\section{Introduction} 
\label{sec:intro}
Blazars, including BL Lacerate (BL Lac) objects, 
are a class of active galactic nuclei (AGN), 
from which a relativistic jet is emanating close to our line of sight. 
In addition to their non-thermal radiation 
ranging from the radio to gamma-ray frequencies and 
strong polarization at radio and optical frequencies, 
one of the most outstanding characteristics of blazars 
is their rapid and high-amplitude intensity variations or flares,  
which provide clues to the flow dynamics, as well as to particle
acceleration and cooling processes operating in the jet.
Since the Ginga (e.g., \cite{PKS2155_Ginga,PKS0548_GINGA})
and ASCA era (e.g., \cite{Mrk421_ASCA_long,HBL_timescale}),
X-ray observations have been one of the most useful tools 
to study the variability of blazars.
In addition, remarkable progress was successively 
accomplished by X-ray observations with
RXTE and Swift observations (e.g., \cite{TeV_Blazars_ASCA,Mrk421_Swift}).

Located at the redshift of $z = 0.031$, 
Mrk 421 is a high energy peaked BL Lac object (HBLs; \cite{HBL_def}). 
The object is one of the brightest extragalactic source 
at the very high energy (VHE) gamma-rays above $\sim 100 $ GeV
\citep{Mrk421_as_TeV}.
A number of studies (e.g., \cite{Mrk421_Param}) revealed that 
the multi-frequency spectral energy distribution (SED) 
of the source and its variation 
are basically well understood within the framework of 
a simple one-zone synchrotron-self-Compton (SSC) model (\cite{SSC}).
For Mrk 421, the X-ray and VHE gamma-ray bands correspond to,
or are located slightly above the peak frequencies 
of the synchrotron and inverse Compton (IC) spectral components, 
respectively.
As a result, 
Mrk 421 is one of the most extensively studied blazars in the X-ray band  
(e.g., \cite{Mrk421_ASCA_long,TeVBlazar_spec_evolve}),
and X-ray flares were frequently reported 
(e.g., \cite{Mrk421_Swift,Mrk421_2008flare}). 

As the first astronomical mission on the Exposed Facility 
of the Japanese Experiment Module ``Kibo'', 
attached to the International Space Station (ISS), 
Monitor of All-sky X-ray Image (MAXI; \cite{MAXI})
started its operation in 2009 August. 
Thanks to its unprecedentedly high sensitivity as an all-sky X-ray monitor, 
and to its capability of real-time data transfer, 
MAXI is able not only to make a continuous monitor of X-ray sources
including AGN,
but also to alert quickly various transient X-ray phenomena,
like flares of blazars. 
Actually, in the 7 month MAXI observation, 
we have successfully made quick alerts of two strong X-ray flares 
from Mrk 421 \citep{Mrk421_flare1,Mrk421_flare2};
in one of these flares the object exhibited the highest X-ray flux 
among those ever recorded. 
In the present paper, we report the X-ray features of Mrk 421 in these flares,
and discuss their implication.

\section{Observation and Results} 
\label{sec:results}
Among the two X-ray instrument onboard MAXI, 
the Gas Slit Camera (GSC) 
and the Solid-state Slit Camera (\cite{MAXI_SSC}),
we analyze only the GSC data in the present paper, 
because of its higher sensitivity and sky coverage ($\sim 97$\% per day). 
The GSC utilizes 12 position-sensitive large-area 
proportional counters sensitive to X-ray photons in 2 -- 20 keV,
although 4 out of them were switched off,  
on 2009 September 22, due to unexpected discharge events.

The MAXI GSC signals from Mrk 421 were integrated orbit by orbit 
within a \timeform{3D} $\times$ \timeform{3D} square
aligned to the scan direction centered on the source, 
while the background level was evaluated 
from two squares with offsets of $\pm$\timeform{3D} 
toward the scan direction in the sky, 
of which the size is same as that of the source region.
We normalized the background-subtracted signal counts 
by a time-integrated effective area of the collimator plus slits
above the proportional counters. 
In the following,
the data from all the activated counters are summed up.

As assistance for the interpretation of the MAXI GSC data, 
we utilized the hard X-ray data 
taken with the Burst Alert Telescope (BAT; \cite{BAT}) onboard Swift. 
The Swift BAT is composed of the coded aperture mask and 
large area ($5200$ cm$^2$) CdZnTe detectors.  
Thanks to its wide field of view ($\sim 1.4$ str), 
the typical sky coverage of the instrument 
exceeds $\sim 70$ \% \citep{BAT_SkyCoverage} 
in one day observation.
The daily Swift BAT lightcurves of Mrk 421 in the $15$ -- $50$ keV range 
are available from the Swift BAT transient monitor results
\footnote{http://heasarc.gsfc.nasa.gov/docs/swift/results/transients/},  
provided by the Swift BAT team.

Figure \ref{fig:lightcurve} (a) and (b) show 
the normalized signal count rates from Mrk 421 
derived with the MAXI GSC in 2 -- 4 keV and 4 -- 10 keV, 
$F_{\rm 2 - 4}$ and $F_{\rm 4 - 10}$ respectively, 
from 2009 August 22 (MJD = 55065) to 2010 March 20 (MJD = 55275).
On the plot, 10 \% of the count rate from the Crab Nebula
(100 mCrab, $F_{\rm 2 - 4} = 0.13$ counts cm$^{-2}$ s$^{-1}$ and
$F_{\rm 4 - 10} = 0.125$ counts cm$^{-2}$ s$^{-1}$; \cite{MAXI_XTEJ1752}) 
is indicated by the dotted lines. 
We notice some relatively long data gaps, 
such as on 2009 September 3 -- 7 (MJD = 55077 -- 55081),
November 12 -- 16 (MJD = 55147 -- 55151),
and 2010 January 19 -- 26 (MJD = 55215 -- 55222). 
During these periods, Mrk 421 was located in the direction 
toward the ISS orbital pole unobservable with the MAXI GSC.
The Swift BAT lightcurve in the 15 -- 50 keV range $F_{\rm 15 - 50}$
is compared in figure \ref{fig:lightcurve} (c). 

The lightcurve 
reveals that Mrk 421 is highly variable in all the energy ranges. 
The fractional root mean square 
variability parameter (e.g., \cite{variablilty_param}) 
in $F_{\rm 2 - 4}$, $F_{\rm 4 - 10}$ and $F_{\rm 15 - 50}$
were evaluated as $0.52$, $0.71$ and $0.70$, 
respectively, from the daily averaged lightcurves. 
The GSC has detected at least four flares 
with an X-ray flux of $\gtrsim 100$ mCrab,  
on 2009 November 1  (MJD = 55136), 
2009 November 11 (MJD = 55146), 
2010 January 1 (MJD = 55197; \cite{Mrk421_flare1}), 
and February 16 (MJD = 55243; \cite{Mrk421_flare2}).
Unfortunately, the observational conditions 
for the first and second flares were found to be relatively bad, 
although the second flare appears to be concurrent 
with the hard X-ray one, detected with the Swift BAT 
on 2009 November 10 (MJD = 55145; \cite{Mrk421_2009_SwiftBAT}).
Therefore, 
we focus on the GSC data associated the third and forth flares 
which are indicated by two arrows, 
denoted as {\bf Epoch A} and {\bf Epoch B} respectively 
in figure \ref{fig:lightcurve} (b). 
During Epoch A, no significant increase in $F_{\rm 15 - 50}$ 
was apparently found, 
while the object brightened in all the energy ranges during Epoch B. 
The flare of Epoch A (MJD = 55197) was not covered 
with the Swift BAT, due to sparse sampling toward Mrk 421 during this period.
In addition, the hardness variation 
between Epoch A and Epoch B (before the first peak) 
was found to be statistically insignificant, as we show below.

Figure \ref{fig:flare} shows 
the 2 -- 10 keV GSC lightcurve, $F_{\rm 2 - 4} + F_{\rm 4 - 10} $,  
during Epochs A and B, together with hardness ratios,
defined as $HR1 = F_{\rm 4 - 10} / F_{\rm 2 - 4}$ 
and $HR2 = F_{\rm 15 - 50} / F_{\rm 2 - 4}$. 
The lightcurve in Epoch A peaks at MJD = 55197.4,
with a 2 -- 10 keV flux of $120 \pm 10$ mCrab, 
averaged over 6 hours around the peak. 
Here and hereafter, the error represents the $1 \sigma$ statistical one. 
The doubling time scales in the rise and decay phases of this flare
were estimated as $t_{\rm r} \lesssim 2 \times 10^{4}$ s and 
$t_{\rm d} \sim 2.5 \times 10^{4}$ s, respectively.
During Epoch B, the source activity continued for about one month,
and we found several flares in the lightcurve.
The peak 2 -- 10 keV flux of the first flare in Epoch B
was measured at MJD = 55243.6 as $164 \pm 17$ mCrab on 6 hour average.  
The decay time of this flare is derived 
as $t_{\rm d} \sim 1.4 \times 10^{5}$ s,
a factor of $5$ -- $6$ longer than that of the Epoch A flare.  
Due to the data gap with $\gtrsim 1 $ day 
just before the peak seen in figure \ref{fig:flare},
we put only a loose upper limit on the rise time as 
$t_{\rm r} < 1.3 \times 10^{5}$ s. 
The lightcurve exhibited the second peak at MJD = 55249.6 
with a 2 -- 10 keV flux of $108 \pm 7 $ mCrab. 

Utilising the daily averaged hardness ratio, 
we investigated quantitatively the spectral variation of Mrk 421 
during Epochs A and B, as summarised in table \ref{table:HR}. 
The hardness of the source was found to stay unchanged 
during the Epoch A within the statistical uncertainties 
($\chi^2/{\rm dof} =  7.3 / 18$ and  $14.9/15$ 
for $HR1$ and $HR2$, respectively).
The average and standard deviation of $HR1$ were derived as 
$\overline{HR1} = 0.74$ and $\sigma_1 = 0.10$, respectively, 
while those of $HR2$ were calculated as  
$\overline{HR2} = 5.3 \times 10^{-2}$, 
and $\sigma_2 = 2.3 \times 10^{-2}$. 
Before the decay of the first flare in Epoch B (MJD = $55232$ -- $55245$; 
corresponding to the arrow denoted as "Before" in figure \ref{fig:flare}),
the object exhibited a hardness similar to that in Epoch A.
No statistically significant spectral variation was observed 
during this phase 
($\chi^2/{\rm dof}= 6.6 / 11$ and  $ 7.4/9$ for $HR1$ and $HR2$).
After the decay of the Epoch B first flare
(MJD = $55246$ -- $55275$; 
indicated by the arrow with "After" in figure \ref{fig:flare}),
the spectrum of the source seems to harden into 
$\overline{HR1} = 0.90 $ and $\overline{HR2} = 10.0 \times 10^{-2}$,
with the standard deviation of $\sigma_1 = 0.15$ and 
$\sigma_2 = 2.7 \times 10^{-2}$, respectively. 
In addition, the source hardness is found to be variable 
($\chi^2/{\rm dof}= 42.1 / 27$ and  $73.8 / 27$ for $HR1$ and $HR2$) 
during this period. 
Toward the end of Epoch B (MJD $\gtrsim 55264$), 
the source appears to recover 
a hardness similar to those before the first peak 
(and hence those in Epoch A). 
In order to visualize these spectral variation associated with the flares,
$HR1$ and $HR2$ are plotted against the source X-ray intensity, 
$F_{\rm 2-4} + F_{\rm 4-10}$, in figures \ref{fig:hardness}. 

Figure \ref{fig:color} shoes 
the color-color plot between $HR1$ and $HR2$ 
with a time resolution of 3 days. 
From this figure, 
we roughly evaluated the shape of the X-ray spectrum of Mrk 421.
If a simple power-law (PL) model was assumed, 
the observed values of $HR1$ correspond to 
a 2 -- 10 keV photon index of $\Gamma = 2$ -- $2.5$,
as shown by the vertical dashed lines in figure \ref{fig:color}.
The systematic uncertainty in the photon index 
due to the current GSC response
is estimated as $\Delta\Gamma = 0.1$ \citep{MAXI_XTEJ1752}.
The figure also indicates a spectral softening from the MAXI GSC
to the Swift BAT energy ranges.

\section{Discussion}  
\label{sec:discussion}
\begin{figure}[h]
\begin{center}
\FigureFile(80mm,80mm){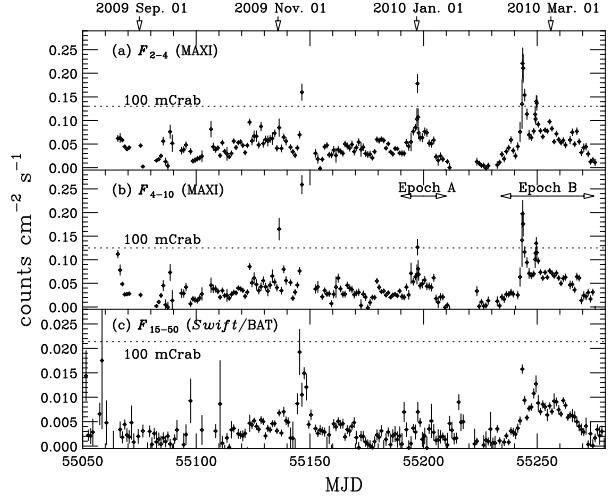}
\end{center}
\caption{Daily lightcurve of Mrk 421. 
The panels (a) and (b) show the normalized count rate
in the 2 -- 4 keV ($F_{\rm 4 - 10}$) 
and 4 -- 10 keV ($F_{\rm 2 - 4}$) ranges, respectively,
measured with the MAXI GSC 
from 2009 August 22 (MJD=55062) to 2010 March 20 (MJD=55275). 
For MJD = 55197, 55243 and 55249, 
$F_{\rm 4 - 10}$ and $F_{\rm 2 - 4}$ are plotted 
in a time resolution of $6$ hours, 
in order to resolve the peak.
The Swift BAT lightcurve in 15 -- 50 keV, $F_{\rm 15 - 50}$, 
is plotted in the panel (c).
All the errors represent the $1 \sigma$ statistical one. 
The dotted lines indicate the count rate of 100 mCrab
in the individual energy ranges. 
The two arrows in the center panel define Epochs A and B. }
\label{fig:lightcurve}
\end{figure}

\begin{figure*}[h]
\begin{center}
\FigureFile(160mm,160mm){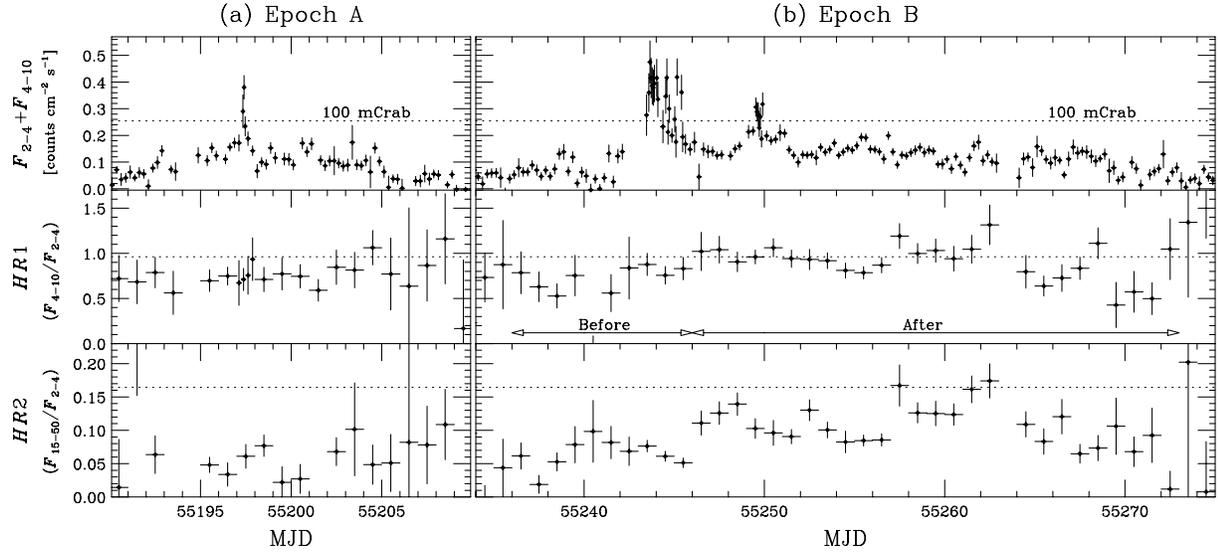}
\end{center}
\caption{Lightcurve of Mrk 421 during Epochs A (panel a) and B (panel b). 
The GSC 2 -- 10 keV count rate 
$F_{\rm 4 - 10} + F_{\rm 2 - 4}$ in a time resolution of 6 hours
is shown in the top.
A time resolution of 1.5 hour is adopted,
when the 6 hour averaged flux is higher than the 100 mCrab level 
(the dotted line). 
The center and bottom display the daily-averaged hardness ratios,
defined as $HR1 = F_{\rm 4 - 10} / F_{\rm 2 - 4}$ and 
$HR2 = F_{\rm 15 - 50} / F_{\rm 2 - 4}$, respectively.
For MJD = 55197 in panel (a), $HR1$ is displayed every 6 hours.  
The hardness for the Crab-like spectrum with 
the photon index of $\Gamma = 2.08$ \citep{crab} 
with the Galactic absorption column density of 
$N_{\rm H} = 3.2\times21 $ cm$^{-2}$ \citep{NH} 
is expressed by the dotted lines.
The long and short arrows in the center of panel (b) define 
the phases before and after the decay 
of the Epoch B first peak (MJD = 55246),
adopted in figures \ref{fig:hardness} and \ref{fig:color}.}
\label{fig:flare}
\end{figure*}

As shown in figures \ref{fig:lightcurve} and \ref{fig:flare},
at least the two active phases of Mrk 421 in 2010 January (Epoch A) 
and February (Epoch B) were extensively monitored with the MAXI GSC.  
The maximum 2 -- 10 keV flux of the flare in Epoch A was derived
as $120 \pm 10$ mCrab at MJD = 55197.4. 
During Epoch B, 
the MAXI GSC revealed a significant long-term activity of the source 
for nearly a month with a 2 -- 10 keV flux of $\gtrsim 50$ mCrab
exhibiting multiple flares.
This demonstrates a great advantage of the MAXI GSC 
for monitoring long-term variations of blazars. 
Especially, the first flare of Epoch B,
of which the peak flux was measured as $164 \pm 17$ mCrab (MJD = 55243.6),
is probably associated 
with a strong and variable activity in the VHE gamma-rays
with the maximum VHE flux exceeding the 10 Crab level,
detected on 2010 February 17 (MJD = 55244) 
by the VERITAS Observatory \citep{Mrk421_VERITAS}.

Since it was identified 
as the first extragalactic VHE $\gamma$-ray source in 1990s
\citep{Mrk421_as_TeV}, 
Mrk 421 has been one of the most extensively studied blazars. 
As a result, a number of X-ray observations of the object 
were conducted with various X-ray observatories, 
including Ginga \citep{Mrk421_Ginga}, 
ASCA \citep{Mrk421_ASCA,Mrk421_ASCA_long,HBL_timescale}, 
RXTE (e.g., \cite{TeV_Blazars_ASCA}), 
XMM-Newton (e.g., \cite{Mrk421_XMM}), 
Suzaku \citep{Mrk421_Suzaku} and so forth. 
Except for in significant X-ray flares, 
its 2 -- 10 keV X-ray flux was reported to be typically below $\sim 50$ mCrab.
Thanks to recent X-ray monitoring observations 
with Swift and the All Sky Monitor onboard RXTE,
strong X-ray flares were successively reported from the object; 
these include the flare in 2006 June with 
the 2 -- 10 keV flux of $\sim 85$ mCrab \citep{Mrk421_Swift},
the flare in 2008 June with $\sim 130$ mCrab \citep{Mrk421_2008flare}  
and the flare in 2009 December with $\sim 100$ mCrab \citep{Mrk421_2009Nov}.
It is important to note that 
the maximum X-ray flux in the first flare of Epoch B, 
measured with the MAXI GSC,
is higher than any other previous results.
Therefore, we have concluded that 
the MAXI GSC discovered the strongest X-ray from Mrk 421.

The color-color plot in figure \ref{fig:color} suggests that 
the MAXI GSC spectrum of Mrk 421 is roughly approximated by a PL 
with a photon index of $\Gamma = 2$ -- $2.5$.
A convex curvature in the MAXI GSC and Swift BAT energy range 
inferred from a comparison of $HR1$ to $HR2$ in figure \ref{fig:color}, 
together with the previous determination of the synchrotron
peak energy in Mrk 421 around the observed photon energies of 
$E_{\rm p } = 0.3$ -- $5$ keV 
(e.g., \cite{TeVBlazar_spec_evolve,Mrk421_XMM,Mrk421_Suzaku}), 
implies a smoothly curved X-ray spectrum of the source.

In the case of HBLs,
where the synchrotron peak frequency is located at the X-ray band, 
the timescale of the X-ray flux variation is thought to be controlled 
mainly by interplay between the following timescales \citep{JK_Thesis}; 
the acceleration timescale of high energy electrons $\tau_{\rm acc}$, 
the radiative cooling time $\tau_{\rm cool}$, 
and the light crossing time of the emission region $\tau_{\rm crs}$.
We, here, describe all the timescales in the observer's frame,
after converting those in the source rest frame, $\tau'$,  
as $ \tau = \tau'/\delta$,
in order to relate them directly to the observed timescales.  
For a significant flare to take place, 
a condition of $\tau_{\rm acc} < \tau_{\rm cool}$ is required.
Assuming a flare is a single event, 
the rise timescale $t_{\rm r}$ is expected to be determined 
by the longer of $\tau_{\rm acc}$ and $\tau_{\rm crs}$, 
while the longer of $\tau_{\rm cool}$ and $\tau_{\rm crs}$ 
is thought to dominate the decay timescale $t_{\rm d}$. 
The soft X-ray lag, 
discovered with ASCA from Mrk 421 in the decay phase of 
an X-ray flare \citep{Mrk421_ASCA},
was regarded to be produced by the energy dependence of $t_{\rm cool}$.
In contrast, symmetric X-ray lightcurves with $t_{\rm r} \sim t_{\rm d}$, 
observed in day-scale flares from several HBLs,
were interpreted to be shaped by $\tau_{\rm crs}$ \citep{HBL_timescale}. 

Because the 2 -- 10 keV MAXI GSC spectrum around the maximum of the flares 
was found to be consistent with $\Gamma \sim 2$,
we regard that the MAXI GSC observed 
near the synchrotron peak frequencies.
Therefore, we adopt the above argument on the flare timescale
to evaluate the physical parameters of Mrk 421 in these flares.
In the flare of Epoch A and the first flare of Epoch B, 
$t_{\rm d}$ was measured to be longer than $t_{\rm r}$. 
This indicates that $t_{\rm r}$ and $t_{\rm d}$ are dominated 
by the different timescales with each other.
As a result, 
we speculate a condition of 
$\tau_{\rm cool} \gtrsim \tau_{\rm crs} \gtrsim \tau_{\rm acc}$ 
or
$\tau_{\rm cool} \gtrsim \tau_{\rm acc} \gtrsim \tau_{\rm crs}$ 
in these flares, 
corresponding to 
$t_{\rm d} \sim \tau_{\rm cool}$ and $t_{\rm r} \gtrsim \tau_{\rm crs}$.

The synchrotron radiation usually dominates other cooling process 
in HBLs, including Mrk 421 (e.g., \cite{JK_Thesis}). 
Assuming an isotropically distributed electrons 
in a randomly oriented magnetic field,
the synchrotron cooling time scale is expressed as  
$\tau_{\rm cool} \sim 1.5 \times 10^{3} B^{-3/2} 
E_{\rm keV}^{-1/2} \delta^{-1/2}$ s 
(re-evaluated from \cite{Mrk421_ASCA}),
with $B$, $E_{\rm keV}$, and $\delta$ 
being the magnetic field in G, the synchrotron photon energy in keV, 
and the beaming factor of the jet, respectively. 
We roughly evaluated a time-averaged magnetic field of 
$B \sim 4.5 \times 10^{-2} (\delta/10)^{-1/3}$ G 
in the flare of  Epoch A,
from the relation of $t_{\rm d} \sim \tau_{\rm cool}$.
Here, we assume the typical synchrotron photon energy 
of $E_{\rm keV} \sim 4$ keV,
by averaging the X-ray spectrum of $\Gamma \sim 2$ 
in the MAXI GSC energy range ($2$ -- $10$ keV).
This is barely consistent with the value determined 
from the one-zone SSC modeling to 
the observed SED of Mrk 421 in previous observations
within the systematic uncertainty
($ B = 0.036$ -- $0.44$ G; \cite{Mrk421_Param}). 
A spectral softening or soft lag, 
due to the energy dependence of $\tau_{\rm cool}$, 
was unobservable within the MAXI GSC signal statistics.
On the other hand,  
we can put an upper limit on the size $R$ of the emission region, 
as $R \le c \tau_{\rm crs} \delta / (1 + z)$, where $c$ is the speed of light. 
The condition of $t_{\rm r} \gtrsim \tau_{\rm crs}$ gives 
$ R \lesssim 6 \times 10^{15} (\delta/10)$ cm. 
This upper limit appears to be consistent with the values 
reported in the previous strong X-ray flares (e.g., \cite{Mrk421_2008flare}). 

\begin{figure*}[h]
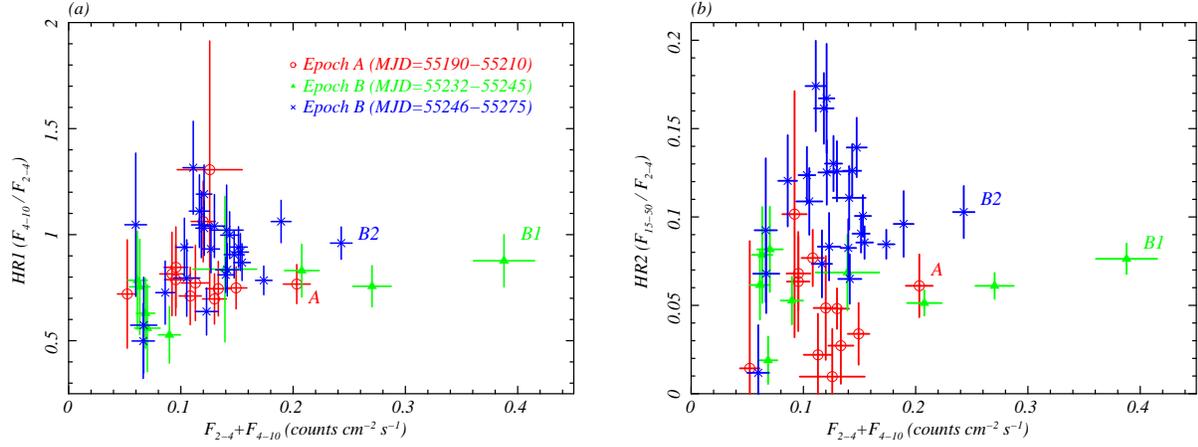

\begin{center}
\includegraphics[angle=-90,width=7.5cm]{figure3a.ps}
\hspace{0.5cm}
\includegraphics[angle=-90,width=7.5cm]{figure3b.ps}
\end{center}
\caption{
$HR1$ (a) and $HR2$ (b) 
plotted against $F_{\rm 2-4} + F_{\rm 4 - 10}$,
in a time resolution of a day.  
Only the data with 
$F_{\rm 2-4} + F_{\rm 4 - 10} > 0.05$ counts cm$^{-2}$ s$^{-1}$ are displayed. 
The error bars indicate the $1 \sigma$ statistical errors.
The red circles show the data during Epoch A,
while the green triangle and blue crosses represent 
the Epoch B data before and after MJD = 55246, respectively 
(indicated the two arrows in the center of figure \ref{fig:flare}).
The data points covering the peaks of the Epoch A flare, 
the first and second flares of Epoch B 
are respectively indicated with A1, B1 and B2, respectively.} 
\label{fig:hardness}
\end{figure*}

\begin{figure}[h]
\begin{center}
\includegraphics[angle=-90,width=8cm]{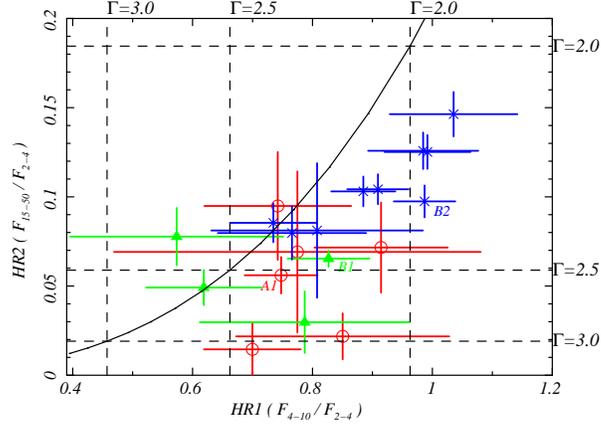}
\end{center}
\caption{Color-color plot between $HR1$ and $HR2$, averaged over 3 days.
The symbol notation same as figure \ref{fig:hardness} is adopted
for the individual phases.
Those covering the peaks of the Epoch A flare,  
the first and second flares of Epoch B 
are indicated with A1, B1 and B2, respectively. 
The error bars indicate the $1 \sigma$ statistical errors.
The dashed lines show the values of $HR1$ and $HR2$, 
for the photon index of $\Gamma = 2.0$, $2.5$ and $3.0$,
assuming that the Galactic absorption column density of 
$N_{\rm H} = 1.9\times20 $ cm$^{-2}$ toward Mrk 421 \citep{NH}. 
If the spectrum is a straight PL in the 2 -- 50 keV range, 
the data points lie on the solid line. }
\label{fig:color}
\end{figure}

In the similar manner,
we put a loose upper limit from $t_{\rm r}$ 
on the region size of the Epoch B first flare 
as $ R \lesssim 4 \times 10^{16} (\delta/10)$ cm.
The magnetic field, 
$B \sim 1.5 \times 10^{-2} (\delta/10)^{-1/3}$ G,
is found to be weaker by a factor of $\sim 3$
than that in the Epoch A flare 
due to $t_{\rm d}$ longer by a factor of $5$ -- $6$,
while the spectral softening 
(from $HR2 = (7.6 \pm 0.9) \times 10^{-2}$ at MJD = 55243 to 
$(5.1 \pm 0.7) \times 10^{-2}$ at MJD = 55245), 
suggested in figures \ref{fig:flare} and \ref{fig:hardness} 
during the decay phase of the Epoch B first flare, 
appears to be related to the cooling process. 
We cannot deny a simple explanation for the possible weak magnetic field 
that this flare was produced in a part of the jet 
different from that of the Epoch A flare
with slightly different physical parameters and/or conditions,
from the MAXI data alone.  
However, we regard that 
$t_{\rm d}$ is not actually determined by $\tau_{\rm cool}$ 
in the case of this flare,
because the derived magnetic field is found to 
be out of range of the typical value of Mrk 421 
in the previous studies (e.g., \cite{Mrk421_Param}). 
Based on close examination on X-ray lightcurve of HBLs 
obtained in the long-look ASCA and RXTE observation 
with a duration of $\gtrsim 1$ week, 
\citet{HBL_timescale} proposed 
a possibility that day-scale flares of HBLs are composed of 
short events with a timescale of $\sim 10^{4}$ s. 
A similar scenario is possible to account for 
the duration of the first flare of Epoch B ($t_{\rm d} \sim 1.6$ day), 
longer by a factor of $5$ -- $6$ than that of the Epoch A flare. 
It is important to note that 
in order to validate the physical quantities estimated here, 
examination of a multi-wavelength SED of Mrk 421, 
widely covering the synchrotron and IC components, 
and its variation are required, 
although it is beyond the scope of the present paper. 

Within the framework of a simple one-zone SSC model, 
the peak energy and luminosity of the synchrotron component 
is thought to scale as 
$E_{\rm p} \propto B \gamma_{\rm p}^2 \delta $ and 
$L_{\rm p} \propto B^{2} 
\gamma_{\rm p}^{2(2-\Gamma_p)} \delta^{2+\Gamma_p} N_{\rm e} V$,
where $\gamma_{\rm p}$, $\Gamma_{\rm p}(< 2)$, $N_{\rm e}$ and $V$ 
is the maximum Lorentz factor of the radiating electrons, 
the synchrotron photon index below $E_{\rm p}$ 
(assuming a simple PL spectrum), 
the normalization of the electron number density spectrum, 
and the volume of the emission region, respectively.
From these scaling relation, 
a change in $B$, $\gamma_{\rm p}$ or $\delta$ is expected 
to be observed as a positive correlation between $E_{\rm p}$ and $L_{\rm p}$,
while only $L_{\rm p}$ is anticipated to increase 
when $N_{\rm e}$ or $V$ (corresponding to the total electron number)
increase. 
The clear positive $E_{\rm p}$--$L_{\rm p}$ correlation,
widely reported from Mrk 421 
in the previous observations (e.g., \cite{TeVBlazar_spec_evolve,Mrk421_XMM}),
is thought to suggest the change in $B$, $\gamma_{\rm p}$ or $\delta$.

The rise of $E_{\rm p}$ is 
thought to be recognised as a rise in $HR1$ and $HR2$. 
The hardness-intensity relation, shown in figure \ref{fig:hardness},
suggests a slightly complicated spectral variation,
which appears not necessarily to follow 
the $E_{\rm p}$--$L_{\rm p}$ positive correlation,
especially in Epoch B. 
We notice that $HR2$ become larger,
after the end of the first flare in Epoch B,
as clearly shown in figure \ref{fig:flare} and table \ref{table:HR}.
Although at the second flare of Epoch B
the source is fainter by $\sim$ 30 \% than at the first one,
its hardness ($HR2 = (10.3 \pm 1.5) \times 10^{-2} $ at MJD = $55249$)
is higher than that at the first flare
($HR2 = (7.6 \pm 1.5) \times 10^{-2} $ at MJD = $55243$). 
In addition, the hardest X-ray spectrum was realized,
when the source flux was about one third of the maximum
at the first peak ($HR2 = (17.4 \pm 2.5) \times 10^{-2} $ at MJD = $55262$). 
In order to interpret such a spectral behaviour
within a simple one-zone SSC framework, 
fine tuning of the physical parameters is required
(e.g., a combination of the increase in $B$, $\gamma_{\rm p}$ or $\delta$ and 
the decrease in $N_{\rm e}$ or $V$).
Otherwise, a multiple-zone/component SSC model 
(e.g., \cite{Mrk421_Suzaku}) could be preferred. 
For detailed examination, 
it is necessary to investigate the multi-frequency SED and its variation.
\color{black}

Thanks to valuable comments from the anonymous referee, 
the present paper has been significantly improved. 
This research has made use of 
the MAXI data\footnote{http://maxi.riken.jp/top/index.php},
provided by RIKEN, JAXA and the MAXI team.
We acknowledge the support 
from the Ministry of Education, Culture, Sports, 
Science and Technology (MEXT) of Japan,
by the Grant-in-Aid for the Global COE Programs
"The Next Generation of Physics, Spun from Universality and Emergence" 
and "Nanoscience and Quantum Physics".
This research was also supported from the MEXT 
by the Grant-in-Aids 
(19047001, 20041008, 20540230, 20244015 , 21340043, 21740140, 22740120).
We are grateful to Dr. M. Kino and Dr. M. Hayashida 
for their advice and assistance.


\clearpage 
\begin{table}[h]
\caption{Summary of spectral variation of Mrk 421}
\label{table:HR}
\begin{center}
\begin{tabular}{llll}
\hline 
Epoch                                      & A                    & B \footnotemark[$^*$] & B \footnotemark[$^\dagger$]  \\ 
\hline 
$\overline{HR1}$ \footnotemark[$\ddagger$] & $0.74$               & $0.73 $               & $0.90 $ \\
$\sigma_1$ \footnotemark[$\S$]             & $0.10$               & $0.13 $               & $0.15 $\\
$\chi^2/{\rm dof}$                         & $ 7.3 / 18$          & $ 6.6 / 11$           & $42.1 / 27$  \\
$\overline{HR2}$ \footnotemark[$\ddagger$] & $5.3 \times 10^{-2}$ & $6.2 \times 10^{-2}$  & $10.0 \times 10^{-2}$ \\
$\sigma_2$ \footnotemark[$\S$]             & $2.3 \times 10^{-2}$ & $1.1 \times 10^{-2}$  & $2.7 \times 10^{-2}$ \\
$\chi^2/{\rm dof}$                         & $14.9 / 15$          & $ 7.4 /  9$           &  $73.8 / 27$ \\
\hline 
\end{tabular}
\end{center}
\footnotemark[$*$] Before the first flare (MJD = 55232 -- 55245).
\par\noindent
\footnotemark[$\dagger$]  After the first flare (MJD = 55246 -- 55275).
\par\noindent
\footnotemark[$\ddagger$] Average of $HR1$ and $HR2$. 
\par\noindent
\footnotemark[$\S$] Standard deviation of $HR1$ and $HR2$.
\end{table}

\clearpage 

\clearpage 

\clearpage 

\clearpage 

\end{document}